\documentclass[graphicx,superscriptaddress,longbibliography,reprint]{revtex4-1}

\usepackage[version=3]{mhchem} 
\usepackage[T1]{fontenc}
\usepackage{siunitx}
\usepackage{graphicx}

\newcommand*{\citen}[1]{%
  \begingroup
    \romannumeral-`\x 
    \setcitestyle{numbers}%
    \cite{#1}%
  \endgroup   
}

\usepackage[normalem]{ulem}
\useunder{\uline}{\ul}{}
\begin{document}
\author{Robin J. Dolleman}
\email{R.J.Dolleman@tudelft.nl}
\affiliation{Kavli Institute of Nanoscience, Delft University of Technology, Lorentzweg 1, 2628 CJ, Delft, The Netherlands}
\author{David Lloyd}
\affiliation{Department of Mechanical Engineering, Boston University, Boston, Massachusetts 02215 United States}
\author{J. Scott Bunch}
\affiliation{Department of Mechanical Engineering, Boston University, Boston, Massachusetts 02215 United States}
\affiliation{Boston University, Division of Materials Science and Engineering, Brookline, Massachusetts 02446 United States}
\author{Herre S. J. van der Zant}
\affiliation{Kavli Institute of Nanoscience, Delft University of Technology, Lorentzweg 1, 2628 CJ, Delft, The Netherlands}
\author{Peter G. Steeneken}
\affiliation{Kavli Institute of Nanoscience, Delft University of Technology, Lorentzweg 1, 2628 CJ, Delft, The Netherlands}
\affiliation{Department of Precision and Microsystems Engineering, Delft University of Technology, Mekelweg 2, 2628 CD, Delft, The Netherlands}

\title{Transient thermal characterization of suspended monolayer MoS$_2$}

\begin{abstract}
We measure the thermal time constants of suspended single layer molybdenum disulfide drums by their thermomechanical response to a high-frequency modulated laser. From this measurement the thermal diffusivity of single layer MoS$_2$ is found to be \num{1.14e-5} m$^2$/s on average. Using a model for the thermal time constants and a model assuming continuum heat transport, we extract thermal conductivities at room temperature between 10 to 40 W/(m$\cdot$K). Significant device-to-device variation in the thermal diffusivity is observed. Based on statistical analysis we conclude that these variations in thermal diffusivity are caused by microscopic defects that have a large impact on phonon scattering, but do not affect the resonance frequency and damping of the membrane's lowest eigenmode. By combining the experimental thermal diffusivity with literature values of the thermal conductivity, a method is presented to determine the specific heat of suspended 2D materials, which is estimated to be 255 $\pm$ 104 J/(kg$\cdot$K) for single layer MoS$_2$.  
\end{abstract}
\maketitle

\section{Introduction}
The distinct electronic \cite{PhysRevLett.105.136805,splendiani2010emerging,eda2011photoluminescence} and mechanical \cite{castellanos2013single,bertolazzi2011stretching} properties of atomically thin molybdenum disulfide opens up possibilities for novel nanoscale electronic \cite{radisavljevic2011single} and opto-electronic \cite{yin2011single,buscema2015photocurrent,buscema2014effect} devices. The large and tunable Seebeck coefficient of single-layer MoS$_2$ makes this material interesting for on-chip thermopower generation and thermal waste energy harvesting \cite{buscema2013large}. Since the power efficiency of these devices depends on the thermal conductivity, it is of interest to study the transport of heat in single-layer MoS$_2$. Several theoretical works have found values of the thermal conductivity $k$ of single layer MoS$_2$ ranging between $k = 1.35$ up to $83$ W/(m$\cdot$K) \cite{liu2013phonon,cai2014lattice,wei2014phonon,li2013thermal,jiang2013molecular}. By exploiting the temperature-dependent phonon frequency shifts in Raman spectroscopy \cite{lanzillo2013temperature}, several experimental works have measured the thermal conductivity of single-layer MoS$_2$ . Experimental values of $k = 34.5$ and $84$ W/(m$\cdot$K) of exfoliated single layer MoS$_2$ have been reported \cite{yan2014thermal,zhang2015measurement}, while single-layer MoS$_2$ grown by chemical vapor deposition was found to show a significantly lower thermal conductivity of 13.3 W/(m$\cdot$K) \cite{bae2017thickness}.

Here, we thermally characterize suspended single-layer MoS$_2$ drum resonators by measuring their thermal time constants. This was achieved by measuring the frequency-dependent vibration amplitude in response to rapidly varying heat flux delivered by a modulated diode laser, similar to previously reported characterization on single-layer graphene \cite{dolleman2017optomechanics}. Since these are frequency based measurements, the result is to first order independent of the absorbed laser power, which greatly facilitates calibration compared to Raman spectroscopy based methods. Furthermore, the method allows one to study relations between the mechanical and thermal properties of the material. From measurements of the thermal time constant $\tau$, we find the thermal diffusivity of MoS$_2$ to be on average $\num{1.05e-5}$ m$^2$/s for 5 $\mu$m diameter drums and $\num{1.29e-5}$ m$^2$/s for 8 $\mu$m drums. Assuming a specific heat value of 373 J/(kg$\cdot$K), this corresponds to $k = 19.8$ W/(m$\cdot$K) and $k = 24.7$ W/(m$\cdot$K). 

\begin{figure*}
\includegraphics{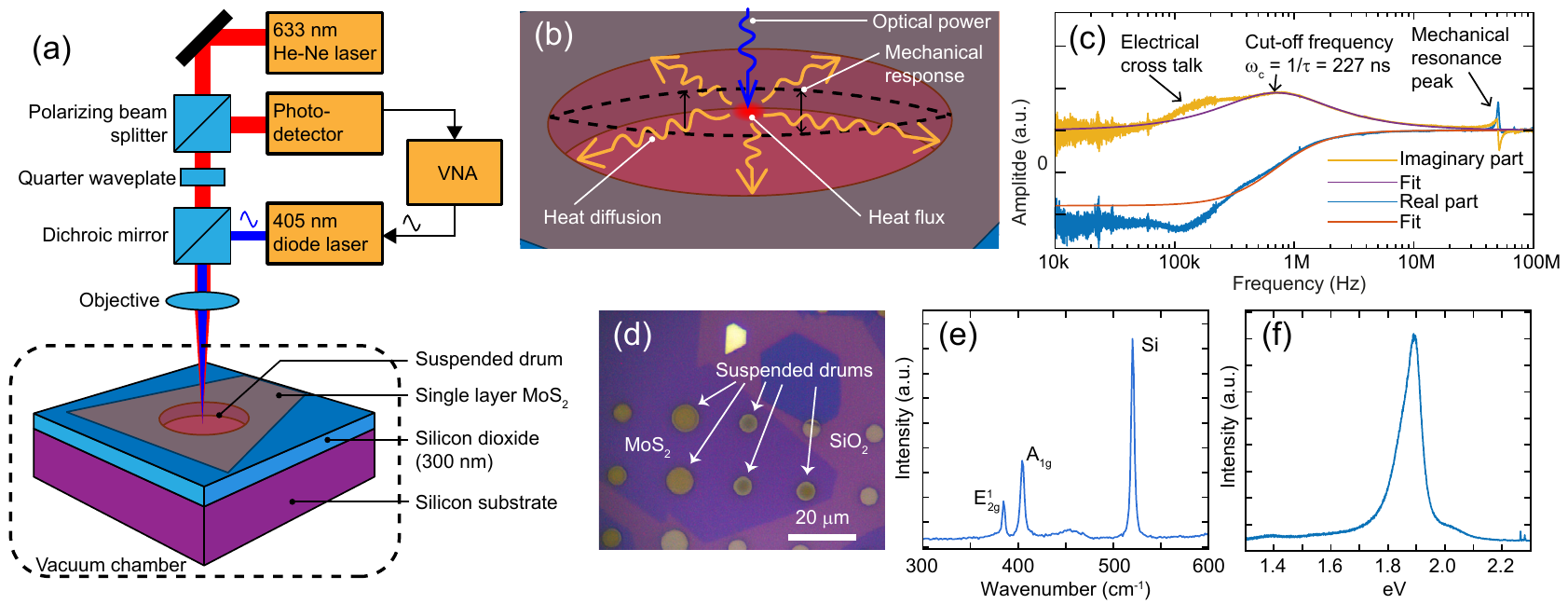}
\caption{(a) Schematic of the samples and the laser interferometer setup to actuate and detect the motion of the single-layer MoS$_2$ drum resonators. (b) Overview of the physical processes involved in measuring the transient properties of heat transport in the drum. (c) Typical experimental result of the real and imaginary part of the amplitude and the fits to find the thermal time constant. At lower frequencies a feature due to electrical cross talk becomes visible due to the low optical gain during the experiment. At higher frequencies the fundamental resonance is clearly visible. (d) Optical image of the device showing a single-layer MoS$_2$ sheet on top of the substrate and several suspended drums. (e) Raman spectrum of the suspended and supported MoS$_2$. (f) Photoluminescence spectrum of the suspended and supported MoS$_2$. The A$^0$ peak position is found at 1.89 eV. \label{fig:setup}}
\end{figure*}
The remainder of this article is structured as follows. Section \ref{sec:exp} describes the experimental setup, fabrication, actuation and read-out of the motion of the single-layered MoS$_2$ drums. The following section \ref{sec:tau} describes the thermal model of the system and how $\tau$ is extracted from the experiments. Section \ref{sec:results} shows the experimental results of $\tau$ and extracts the value of the thermal diffusivity. This section also examines the relation between mechanics and thermal transport. Section \ref{sec:disc} contains an extensive discussion elaborating on the possible causes of the large device-to-device variation, the specific heat of MoS$_2$ and compares the present results to single-layer graphene. The conclusions of this work are then outlined in section \ref{sec:conc}.

\section{Experimental setup}\label{sec:exp}
We use a substrate with many circular cavities to perform the experiment. The fabrication starts with a silicon chip with 285 nm of silicon dioxide. Circular cavities of approximately 300 nm deep and with a diameter of 8 and 5 micron are etched in the oxide layer. Many single layer MoS$_2$ flakes grown by chemical vapor deposition are transferred over the substrate by a dry transfer method to create suspended drum resonators as drawn in Fig. \ref{fig:setup}(a). An optical image of several devices is shown in Fig. \ref{fig:setup}(d). The Raman and photoluminescence (PL) spectra of both the suspended MoS$_2$ flakes are shown in in Figs. \ref{fig:setup}(e) and (f), data was taken on suspended drums to prevent the effects of substrate doping \cite{sercombe2013optical,scheuschner2014photoluminescence,buscema2014effect}. These measurements ensure that the MoS$_2$ flakes are single-layer, since no indirect transition is observed in the PL spectrum (Fig. \ref{fig:setup}(f)) \cite{scheuschner2014photoluminescence}. In the Raman spectra (Fig. \ref{fig:setup}(e)) the E$^1_{2g}$ peak is found at 384.9 cm$^{-1}$ and the A$_{1g}$ peak at 404.5 cm$^{-1}$, also in accordance with single-layer MoS$_2$ \cite{li2012bulk}.  Furthermore, the positions of both the E$^1_{2g}$ Raman peak and PL A$^0$ (1.89 eV) suggests that no large strains ($>1$\%) are induced by the transfer \cite{lloyd2016band}.  More details on the CVD growth and transfer can be found in ref. \citen{lloyd2017adhesion}. The samples are kept in an atmosphere with a maximum pressure of $\num{1e-6}$ mbar for two weeks before and during the experiment to ensure all gas has escaped from the cavity. 

Figure \ref{fig:setup}(a) also shows a schematic drawing of the interferometer setup used to actuate and read-out the motion of the membrane. The red laser intensity on the photodiode is used to read-out the motion using Fabry-Perot interferometry between the moving membrane and the fixed back-mirror, which is the silicon \cite{castellanos2013single,dolleman2017amplitude,bunch2007electromechanical}. The blue laser heats up the membrane, which causes the membrane to move due to thermal expansion \cite{dolleman2017optomechanics,dolleman2017graphene}. The blue laser is power modulated using the output of a vector network analyzer (VNA). The input of the VNA is connected to the photodiode that detects the reflected red laser intensity. A dichroic mirror is used to prevent the blue laser light from reaching the photodiode. The VNA measures both the amplitude and the phase of the transmitted signal. All parasitic phase shifts in the electrical and optical components are measured by directly pointing the blue laser at the photodetector and are eliminated by using the measured transmission function to deconvolve the experimental results \cite{dolleman2017optomechanics}.

\section{Thermal time constant}\label{sec:tau}
Due to the diffusion of heat through the membrane, there will be a time delay between the optical power delivered to the membrane and the membrane's motion (Fig. \ref{fig:setup}(b)). The diffusion of heat can be described by the heat equation:
\begin{equation}\label{eq:heateq}
\rho c_p \frac{\mathrm{d}T}{\mathrm{d}t} - k \nabla^2 T = P,
 \end{equation}
 where $T(\mathbf{x},t)$ is the temperature and $P(\mathbf{x},t)$ the heat flux applied to the membrane. $\rho$ is the density of the material, $c_p$ the specific heat, $k$ the thermal conductivity, $\mathbf{x}$ is the position vector and $t$ is time. By separation of variables, and by using a lumped element model with incident laser heat flux $P = P_{\mathrm{ac}}e^{i \omega t}$, Eq. \ref{eq:heateq} can be simplified, which results in a single relaxation time approximation for the time-dependent temperature:
\begin{equation}\label{eq:lumped}
  C \frac{\mathrm{d} \Delta T}{\mathrm{d} t}+\frac{1}{R} \Delta T=P_{\mathrm{ac}} e^{i\omega t},
  \end{equation}
 where $C$ is the heat capacity and $R$ the thermal resistance. Below the resonance frequency, the motion $z=z_{\omega} \mathrm{e}^{i\omega t}$ is proportional to the temperature change, $z=A \Delta T$, such that it follows from Eq. \ref{eq:lumped} that \cite{dolleman2017optomechanics,metzger2008optical}:
\begin{equation}
 z_\omega=\frac{A P_{\mathrm{ac}} R}{i \omega \tau +1} = A P_{\mathrm{ac}} R \frac{1-i \omega t}{1+\omega^2 \tau^2}
\label{heateq}
\end{equation}
where $A$ is a proportionality constant that will be obtained by fitting and $\tau = RC$ the thermal time constant of the suspended drum.

The thermal time constant $\tau$ can be determined from the measured thermomechanical frequency response of the drum over several decades using the setup in Fig. \ref{fig:setup}(a). Figure \ref{fig:setup}(c) shows the real and imaginary part of the experimentally obtained frequency response from a MoS$_2$ drum with a diameter of 8 $\mathrm{\mu}$m. It follows from eq. \ref{heateq} that the imaginary part of the response function has a maximum amplitude at $\omega \tau=1$. This maximum is indeed observed at a cut-off frequency of $\omega_c =2 \pi \times 800$ kHz in Fig. \ref{fig:setup}d, which is far below the membrane's lowest resonance frequency such that the relation $z_{\omega}= A \Delta T$ is valid. By fitting the imaginary part using eq. \ref{heateq} the thermal time constant of the membrane is determined to be $\tau=1/\omega_c=227$ ns. The resonance peaks were analyzed by fitting a harmonic oscillator model to the data, from which the resonance frequency and quality factor is found. Although both the real and imaginary part of the response function fit well to equation (3), deviations around 300 kHz are observed which are attributed to electrical cross-talk, most likely due to capacitive coupling to the optical table containing the experimental setup. Because the laser powers are low in these experiments to prevent damage to the drums, the total optical signal on the photodiode is very low, making the system very susceptible to parasitic cross-talk. The low frequency data was excluded for the fit in order to prevent cross-talk from affecting the value of $\tau$.

 \begin{figure*}
\includegraphics{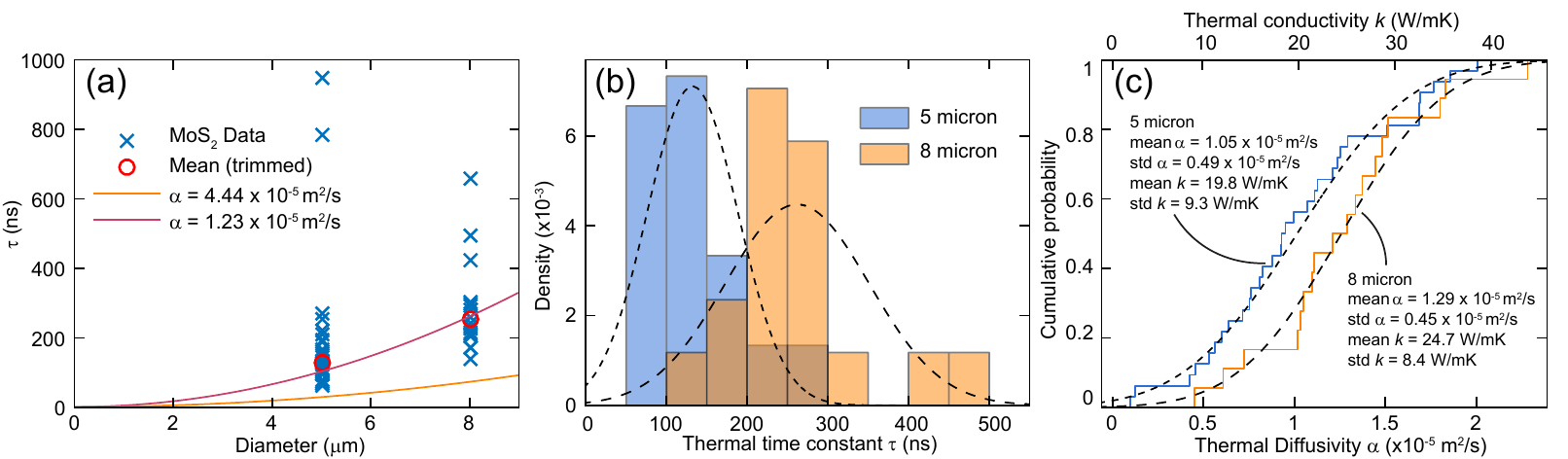}
\caption{(a) Thermal time constants as function of diameter. Predictions using eq. \ref{eq:inttau} are plotted with several values of $k$ obtained from literature: $k = 23.3$ W/(m$\cdot$ K) corresponding to $\alpha = \num{1.23e-5}$ m$^2$/s, \cite{cai2014lattice}   and $84$ W/(m$\cdot$K) to  $\alpha = \num{4.44e-5}$ m$^2$/s. \cite{zhang2015measurement} (b) Density plot of the thermal time constant for both diameter, drums with extremely large values of $\tau$ and low resonance frequency were excluded. (c) CDF of the thermal conductivity $k$ estimated from the values of $\tau$ using  $c_p = 373.5$ J/(kg$\cdot$K) and $\rho = 5060$ kg/m$^3$.  \label{fig:data}}
\end{figure*}
\section{Results}\label{sec:results}
Frequency response fits as shown in Fig. \ref{fig:setup}(c) are obtained on a total of 32 single layer MoS$_2$ drums with a 5 micron diameter and 18 drums with a 8 $\mathrm{\mu}$m diameter. Figure \ref{fig:data}(a) shows the experimentally obtained values from all the drums as function of drum size and Fig. \ref{fig:data}(b) shows a density plot for both diameters. Significant spread in the value of $\tau$ is found, even for drums of the same diameter. To exclude large effects of outliers, we only analyzed 80\% of the samples with value $\tau$ closest to the mean and find $\bar{\tau} = 126$ ns for the 5 micron diameter drums and $\bar{\tau} = 253$ ns for the 8 micron drums. 

\citeauthor{aubin2004radio} derived an expression for the thermal time constant for a uniformly heated circular drum  \cite{aubin2004radio,bunch2008mechanical}:
\begin{equation}\label{eq:inttau}
\tau = \frac{a^2 \rho c_p}{\mu^2 k},
\end{equation}
where $a$ is the drum radius, $\rho $ the density, $c_p$ the specific heat and $k$ the thermal conductivity the material. For a uniformly heated drum, $\mu = 2.4048$ is the first root of the Bessel function $J_0(x)$. However, in the experiments the membrane is heated by a focused laser spot in the center of the drum. We therefore use a numerical COMSOL model that adapts the value of $\mu$ by taking a point heat source in the center of the membrane. The measurement of the temperature is taken as the average temperature over the surface over the drum, since we expect the mechanical response to depend on the temperature field in the entire drum. From the simulations it was found that $\mu^2 = 5.0$ is an accurate representation of the experiments. This should predict the value of $k$ with an error less than 10\% as long as $15 < k < 100$ W/(m$\cdot$K) and assuming that $c_p = 373.5$ J/(kg$\cdot$K) (See Supplemental Information). Using Eq. \ref{eq:inttau} we can estimate the thermal diffusivity of MoS$_2$ $\alpha = k/ \rho c_p$:
\begin{equation}
\alpha = \frac{a^2}{5 \tau}.
\end{equation}
This expression was used to estimate the thermal diffusivity for each drum as shown in Fig. \ref{fig:data}(c). We find the diffusivity is slightly diameter-dependent with an average diffusivity $\bar{\alpha} = \num{1.05e-5}$ m$^2$/s for the 5 micron drums and $\bar{\alpha} = \num{1.29e-5}$ m$^2$/s for the 8 micron drums.

Based on known values of $c_p$ and $\rho$ of molybdenum disulfide at room temperature ($c_p = 373.5$ J/(kg$\cdot$K) and $\rho = 5060$ kg/m$^3$) we can estimate $k = a^2 \rho c_p/(5 \tau)$ from experimental values of $\tau$. Fig. \ref{fig:data}(c) shows the cumulative density function calculated for each drum. We find a mean of $k$, $\bar{k} = 19.8$ W/(m$\cdot$ K) with a standard deviation of $ 9.3$ W/(m$\cdot$ K) for the 5 micron drums and for the 8 $\mu$m drums we find $\bar{k} = 24.7$ W/(m$\cdot$K) with standard deviation $\sigma_k = 8.4$ W/(m$\cdot$ K). We thus observe a considerable spread between devices. Moreover, most of the values of $k$ found here are smaller compared to previous observations in literature that used exfoliated MoS$_2$ devices \cite{yan2014thermal,zhang2015measurement}, but are larger than CVD MoS$_2$ values \cite{bae2017thickness}.

\subsection{Comparison to the resonant properties}
\begin{figure}
\includegraphics{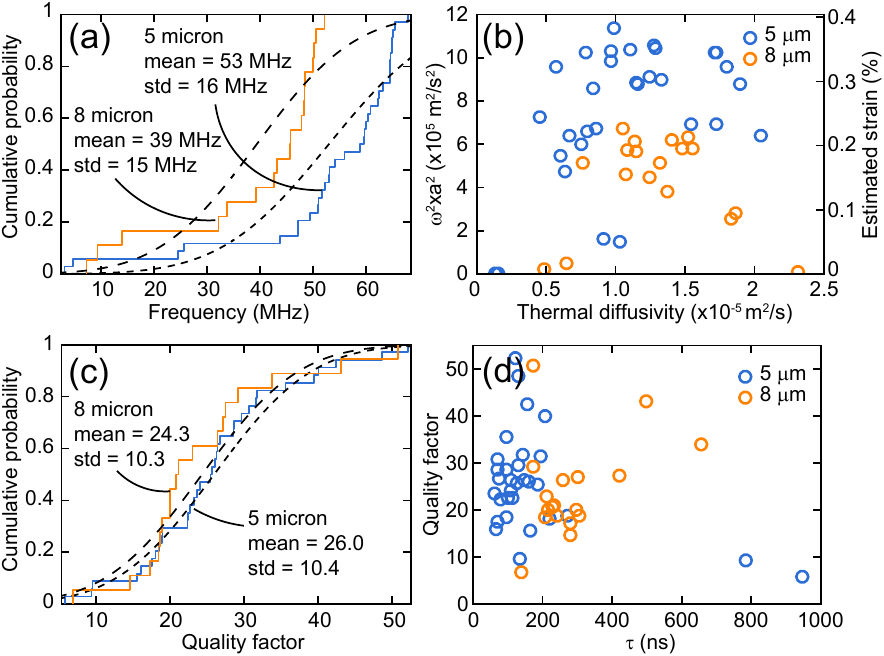}
\caption{Investigation of correlations between the mechanical and thermal properties. (a) CDF of the resonance frequency for both diameters. (b) Scatter plot with the thermal diffusivity on the horizontal axis and frequency times radius squared (which is proportional to tension) on the vertical axis. (c) CDF of the quality factor for both diameters. (d) Scatter plot with the thermal time constant on the horizontal and quality factor of resonance on the vertical axis.  \label{fig:mechdata}}
\end{figure}
The transient mechanical characterization allows one to study whether the mechanical properties of the suspended drums are correlated to the thermal properties. This might be expected, since the acoustic phonon velocities can be tension dependent, which would result in a correlation between the resonance frequency and the thermal diffusivity. Also mechanical damping in graphene due to defects could cause increased phonon scattering, which would lead to a lower thermal conductivity for drums with a low mechanical Q.

  To study this, the resonance peaks were fitted by a harmonic oscillator model to extract the resonance frequency and the quality factor. The distribution of all the resonance frequencies is shown in Fig. \ref{fig:mechdata}(a) and the quality factors are shown in Fig. \ref{fig:mechdata}(c). We first investigate whether the thermal diffusivity is affected by strain in the resonator. The fundamental resonance frequency $f$ of a circular drum resonator is given by:
\begin{equation}
f = \frac{2.4048}{2 \pi a} \sqrt{\frac{n_0}{\rho h}},
\end{equation} 
where $h$ is the thickness of the drum and $n_0$ the tension in the membrane. From this, we deduce that $f^2 a^2 \propto n_0$ if $\rho h$ is the same for each drum. Figure \ref{fig:mechdata}(b) shows a scatter plot of $f^2 a^2$ versus the thermal diffusivity for each drum, the strain was estimated assuming the membrane has the ideal mass and the 2D Young's modulus was taken as 160 N/m \cite{liu2014elastic,lloyd2017adhesion}. No meaningful correlation between tension and the thermal diffusivity could be uncovered. 

We further investigate whether the mechanical dissipation is related to the heat transport properties of these drums by examining the correlations to the quality factor. Figure \ref{fig:mechdata}(d) shows a scatter plot of the quality factor of resonance versus the thermal time constant. No significant correlation between the thermal time constant and the quality factor of resonance is found from the experimental data. The quality factor is nearly independent of diameter as shown in Fig. \ref{fig:mechdata}(c), we find $\bar{Q} = 26.0$ with standard deviation $10.4$ for the 5 $\mu$m drums and $\bar{Q} = 24.3$ with standard deviation $\sigma_Q = 10.3$ for the 8 $\mu$m drums. 

\subsection{Phonon relaxation time and mean free path}
The thermal conductivity can be expressed as $k \approx \rho c_p v \lambda$, \cite{pop2012thermal} where $v$ and $\lambda$ are appropriately averaged phonon group velocity and mean free path, respectively. Substituting this expression in eq. \ref{eq:inttau} gives:
\begin{equation}\label{phonon}
  \tau = \frac{a^2 }{5  v \lambda}= \frac{a^2 }{5  v^2 \tau_{ph}},
\end{equation}
where $\tau_{ph}$ is the phonon relaxation time. We take the averaged velocity as $v \approx 300$ m/s based on calculations from several theoretical works \cite{doi:10.1002/andp.201500354,GAN2016745,cai2014lattice} and use Eq. \ref{phonon} to estimate $\tau_{ph}$ and $\lambda$. For the 5 micron drums we find an average phonon relaxation time and mean free path of 116 ps and 34.9 nm, respectively. For the 8 micron drums we find 143 ps and 43.2 nm. For both cases we again find device-to-device variations due to the spread in the measured values of $\tau$.

\begin{figure}
\includegraphics{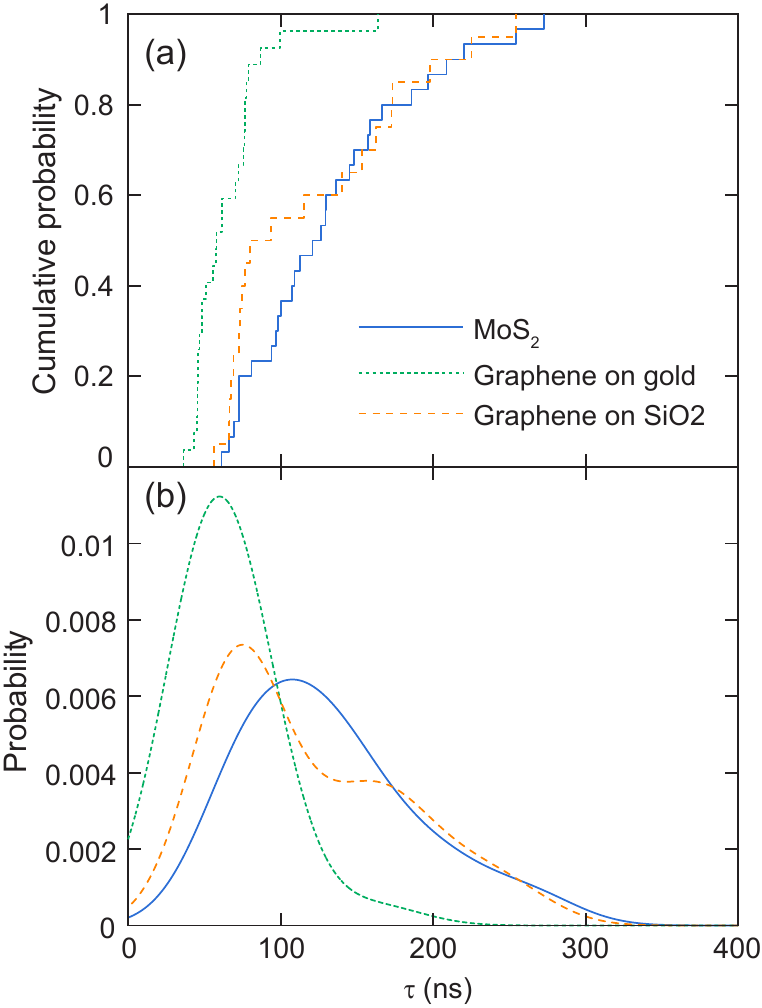}
\caption{(a) Cumulative probabilities from the experimental values of the thermal time constant in this work and for the case of single layer graphene for drums with a 5 micron diameter. (b) Empirical distribution functions found by fitting a Kernel distribution with a 30 ns bandwidth to the data. \label{fig:graphene-mos2}}
\end{figure}
\section{Discussion}\label{sec:disc}
\subsection{Comparison to single-layer graphene}
In Fig. \ref{fig:graphene-mos2} we compare the experimentally obtained values of $\tau$ with experimentally obtained values of single layer graphene (data from previous work in ref. \citen{dolleman2017optomechanics}) for drums with a 5 $\mu$m diameter. From the CDF in Fig. \ref{fig:graphene-mos2}(a) it can be seen that both materials have a thermal time constant with the same order of magnitude. This is striking because even in the worst case scenario (CVD graphene with a lot of defects, $k \approx 600$ W/(m$\cdot$K)) graphene should have a thermal diffusivity at least ten times higher than MoS$_2$. In this previous work on single-layer graphene we attributed the anomalous diameter-dependence of $\tau$ to boundary effects that were limiting the heat transport. Since we only measured two diameters in this work, we cannot use diameter dependence to draw conclusions. Nevertheless, the values of $\tau$ on MoS$_2$ are in good agreement with the theory of diffusive heat transport. This can be seen by comparing the measured values of $\tau$ to the theoretical predictions from literature as shown in Fig. \ref{fig:data}(a). Any effects of a thermal boundary resistance based on the measurements on MoS$_2$ are too small to be discerned. Molybdenum disulfide has a much lower thermal conductivity than graphene, which means that the intrinsic thermal resistance is more important than thermal resistance at the boundary of the drum, if such a resistance is present at all in the case of MoS$_2$.

\subsection{Relation between mechanical and thermal properties}
We could not uncover any meaningful correlation between strain and the thermal diffusivity from the experimental data. The spread in the strain between the devices estimated from the resonance frequency is no more than 0.4\%, which should result in a spread in the thermal conductivity of approximately 3\% \cite{zhu2015thermal}. The measured device-to-device spread is significantly larger and strain-dependence is thus not the cause of the observed variations. It should be considered however that the value of $f^2 a^2$ could actually show spread between devices due to variations in the mass due to polymer contamination. 

\subsection{Device-to-device spread}
The observed device-to-device variations in $\tau$ might be attributed to variations in microscopic (point defects) and macroscopic imperfections between devices, that could alter the phonon relaxation times between devices explaining our result in Fig. \ref{fig:data} (c). From calculations from the literature \cite{doi:10.1002/andp.201500354} using the Boltzmann transport equation for phonons, we would expect a mean free path of 316.5 nm for naturally occurring MoS$_2$. The significantly shorter mean free paths ($\sim$20 to 60 nm) found here might be related to our use of CVD MoS$_2$ rather than pristine exfoliated samples. Additional defects can increase the phonon scattering rate, lowering the phonon relaxation time and the mean free path. This could also explain why our experimentally obtained value of $k$ is smaller than other experimental observations, since previous works employed exfoliated membranes \cite{yan2014thermal,zhang2015measurement}. Most of the drums show a higher value of $k$ than previous observations on CVD-grown MoS$_2$ \cite{bae2017thickness}, which could be related to differences in quality of the sample. The value of the mean free path shows that $\lambda <<  a$, this supports our notion that heat transport can be described by continuum models in these device.

\subsection{Specific heat}
Given the arguments above, the significant spread in $\tau$ is most likely related to the scattering mechanisms. However we cannot fully exclude the possibility that the heat capacity of the drums is responsible for the spread in $\tau$. Little is known about potential mechanisms that can affect the specific heat of single-layered two-dimensional materials due to the lack of experimental data. However, the specific heat is most likely not very different from the bulk material since the number of vibrational degrees of freedom is the same. Also weak temperature dependence of the value of $c_p$ is expected since the experiments are performed above the Debye temperature, therefore most degrees of freedom in the lattice are thermalized.

What we can conclude is that some of the literature values of $k$ are impossible to have occurred in our measurements, since they would violate the Petit-Dulong limit ($c_p = 468.8$ J/(kg$\cdot$K)). The fastest 5 micron diameter drum has $\tau = 61$ ns, which means that there is a limit on the thermal conductivity: $k \leq 48$ W/(m$\cdot$K). For the fastest 8 micron diameter drum, $\tau = 138$ ns and it is impossible that the thermal conductivity of this drum exceeded 55 W/(m$\cdot$K). Therefore, the highest reported value of $k = 84$ W/(m$\cdot$K), \cite{zhang2015measurement} cannot have occurred in the drums used in this study. Also, the reported value of $k = 34.5$ W/(m$\cdot$K), \cite{yan2014thermal} would implicate that the Petit-Dulong limit is violated in most of the devices. 

The most representative study, since it uses both CVD MoS$_2$ and conducted the experiment in vacuum, is $k = 13.3 \pm 1.4$ W/(m$\cdot$K) \cite{bae2017thickness}. Using this value, we can use the experimentally obtained values of $\tau$ to estimate the specific heat of MoS$_2$. For the 5 micron drums, we find $c_p = 278 \pm 118$ J/(kg$\cdot$K) and for the 8 micron drums we find $c_p = 215 \pm 73$ J/(kg$\cdot$K). The errors represent the standard deviation due to the large device-to-device spread, nevertheless this analysis suggests that most of the devices have a specific heat that is significantly lower than the bulk value. 
 Future work can combine the transient characterization with existing methods, such as Raman spectroscopy or electrical heaters, to extract the thermal resistance $R$. In that case the heat capacity $C$ can be derived and provide more accurate measurements on the specific heat of 2D materials. The transient characterization thus provides a means to perform calorimetry on suspended 2D materials. 

\section{Conclusion}\label{sec:conc}
We measured the thermal time constants of suspended monolayer molybdenum disulfide drums. In contrast to previous measurements on single layer graphene, we find that the values of $\tau$ are in agreement with classical Fourier theory of heat transport. From the values of $\tau$ we can estimate the thermal conductivity to be between 10 and 40 W/(m$\cdot$K), which is lower than previous measurements on exfoliated MoS$_2$ but in agreement with measurements on CVD-grown MoS$_2$. Significant device-to-device variation in thermal time constants is observed. This variation is not correlated to the resonance frequency or Q-factor of the membranes, which shows that mechanisms that determine the macroscopic damping are probably not responsible for the observed spread. We therefore conclude that the variations in thermal diffusivity are caused by microscopic defects that have a large impact on phonon scattering, but do not affect the resonance frequency and damping of the membrane's lowest eigenmode. The method can be used to estimate the specific heat of single layer MoS$_2$, with our results suggesting its value might be lower than the bulk value. Future work can combine this technique with existing thermal conductivity measurements to perform calorimetry on suspended 2D materials, enabling one to determine whether the specific heat of 2D materials is equal to its bulk value.
 
\begin{acknowledgements}
This work is part of the research programme Integrated Graphene Pressure Sensors (IGPS) with project number 13307 which is financed by the Netherlands Organisation for Scientific Research (NWO).
The research leading to these results also received funding from the European Union's Horizon 2020 research and innovation programme under grant agreement No 785219 Graphene Flagship. J.S.B. and D.L. were funded by the National Science Foundation (NSF) grant no. 1706322 (CBET: Bioengineering of Channelrhodopsins for Neurophotonic and Nanophotonic Applications) and Boston University.
\end{acknowledgements}

\newpage
\onecolumngrid
\appendix
\section*{Supplemental information}
\subsection*{COMSOL model}
Here we show the COMSOL model used to derive the expression for $\tau$ used in the main section of the paper. An analytic expression was derived for the thermal time constant $\tau$ in the case of a uniformly heated circular disk \cite{aubin2004radio}:
\begin{equation}\label{eq:inttau}
\tau = \frac{a^2 \rho c_p}{\mu^2 k},
\end{equation}
where $a$ is the drum radius, $\rho $ the density, $c_p$ the specific heat, $k$ the thermal conductivity the material and $\mu = 2.4048$ is the first root of the Bessel function $J_0(x)$. Since the experiment uses a laser spot with a size that is much smaller than the drum diameter to heat the drum, equation \ref{eq:inttau} needs to be modified in order to accurately describe the time constant of the system. Our approach is to choose a fixed value of the specific heat $c_p = 373.5$ J/(kg$\cdot$K) and vary both $a$ and $k$ to find a new value of $\mu$ that will enable us to accurately determine the value of $k$ or the thermal diffusivity $\alpha$ from the experiment. 

\begin{figure*}[b]
\includegraphics{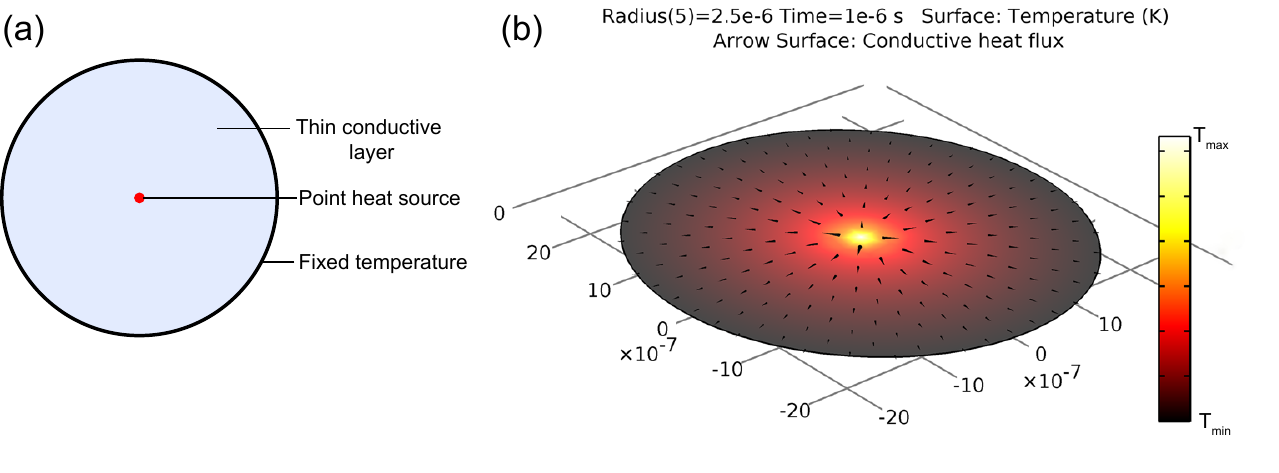}
\caption{(a) Schematic drawing of the simulation. (b) Temperature profile at the end of the simulation for a 5 micron diameter drum. \label{fig:comsolsetup}}
\end{figure*}
Figure \ref{fig:comsolsetup} shows the setup of the COMSOL simulation in order to find the thermal time constant $\tau$. A simple circular domain was defined and the heat transport is simulated using the ``heat transport in thin shells'' module. A point source in the center was used to simulate the heat flux and the boundaries of the domain were kept at a fixed temperature. 

\begin{figure*}
\includegraphics{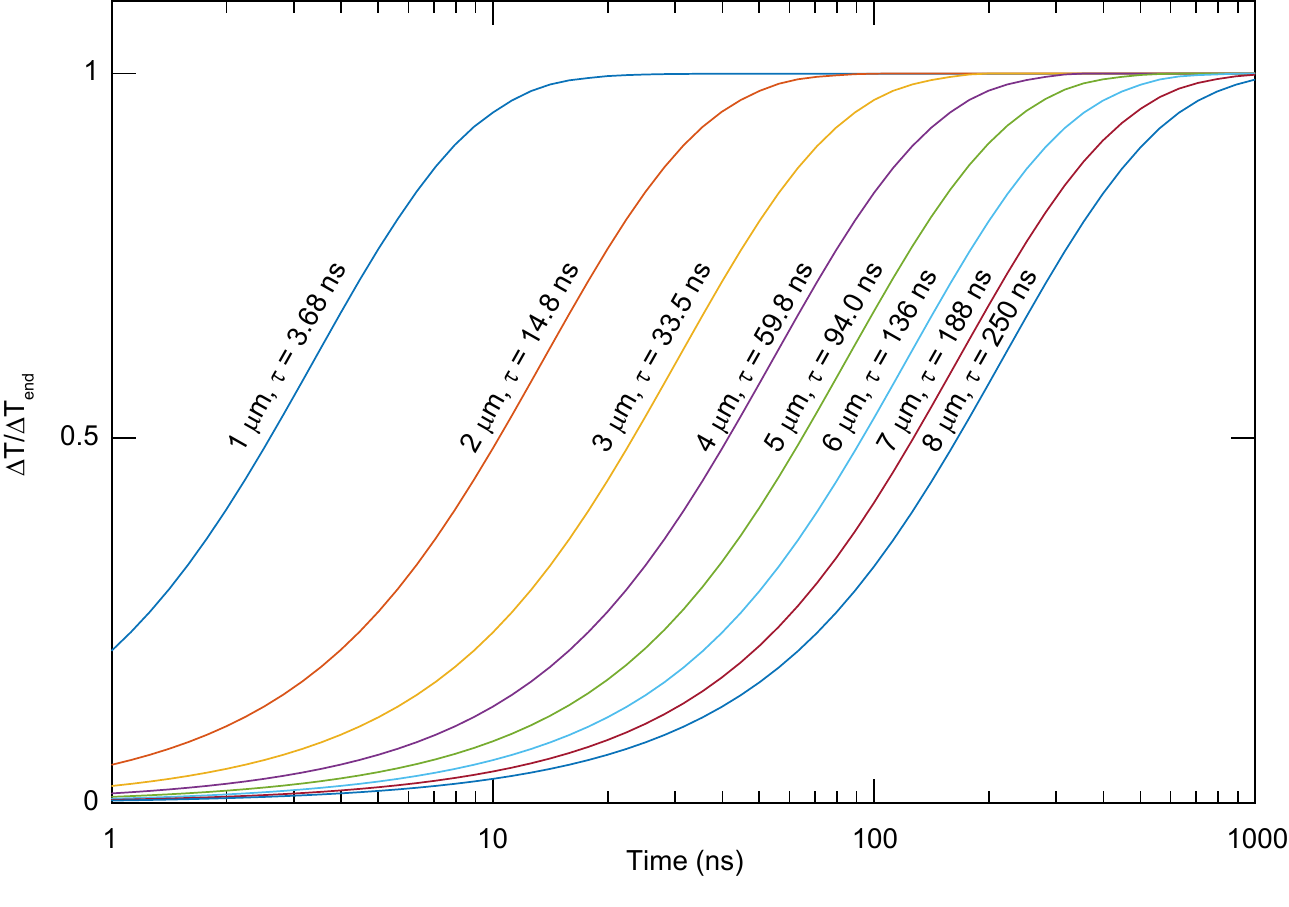}
\caption{Simulated temperature as function of time, from which $\tau$ can be derived for different diameters. The temperature is calculated using the average value over the drum surface. \label{fig:timetraces} }
\end{figure*}
In order to find the time constant $\tau$, a time-dependent simulation was performed that simulates the response to a step function in the heat source. The resulting time dependent temperature increase was calculated by taking the average over the entire domain. This results in the time-dependent traces shown in Fig. \ref{fig:timetraces}. For each trace, the time constant is found by fitting:
\begin{equation}
T(t) = T_0+T_{\mathrm{end}} \exp{(-t/\tau)}.
\end{equation}

\begin{figure*}
\includegraphics{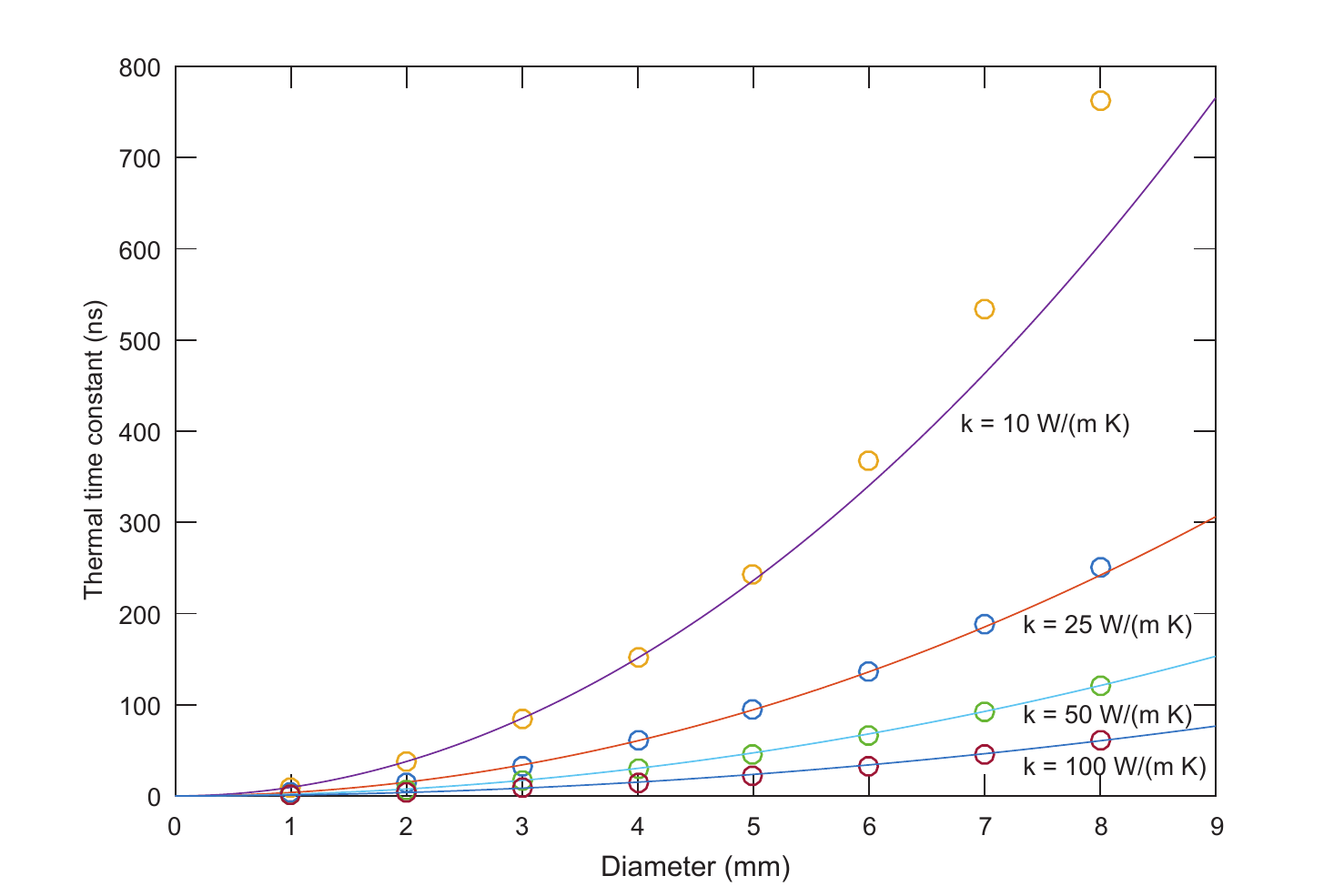}
\caption{Diameter dependence of $\tau$ for different diameters and different values of $k$. \label{fig:model}}
\end{figure*}
The diameter-dependence is simulated using a range of values of $k$, we selected a suitable range by selecting values found in literature. It is found that for the range between $10 < k <100$ we find values of $\mu^2 \approx 5.0$. This model is shown as solid lines in Fig. \ref{fig:model}. We find that $\mu^2 = 5.0$ yields good agreement at the higher values of $k$. Low values of $k$ results in larger deviations at larger diameters, which can also be attributed to the $a^2$ dependence of $\tau$. From this model, we find that the value of $\tau$ with $\mu^2 = 5.0$ should produce the correct value of $k$ or $\alpha$ within 10\% error as long as $15 < k < 100$ W/(m$\cdot$K) and $2a \leq 8$ $\mu$m.

\end{document}